\documentclass[aps,preprint,showpacs,superscriptaddress]{revtex4}

\usepackage{graphicx}
\usepackage{amsmath}
\usepackage{bm}

\newcommand{\be}{\begin{equation}}
\newcommand{\ee}{\end{equation}}
\newcommand{\bel}[1]{\be\label{#1}}
\newcommand{\re}[1]{Eq.~(\ref{#1})}

\newcommand{\ds}{\displaystyle}
\newcommand{\ov}[1]{\overline{#1}}
\newcommand{\psib}{\ov{\psi}}
\newcommand{\hsp}{\hspace*{1pt}}
\newcommand{\intp}{\int\frac{d^3 p}{(2\pi)^3}}
\newcommand{\ENp}{\sqrt{p^2+m^2}}
\newcommand{\dd}{\partial\hspace{-6pt}/}

\begin{document}

\title
{How far is normal nuclear matter\\ 
from the chiral symmetry restoration?~\footnote{
We dedicate this work to Prof. S.T.~Belyaev, the great
theoretical physicist and colleague, on occasion of his
80th birthday}
}
\author{I.N. Mishustin}

\affiliation{The Kurchatov Institute, Russian Research Centre,
123182 Moscow, Russia}

\affiliation{Institut~f\"{u}r Theoretische Physik,
J.W. Goethe Universit\"{a}t,\\
D--60054 Frankfurt am Main, Germany}

\affiliation{The Niels Bohr Institute,
DK--2100 Copenhagen {\O}, Denmark}

\author{L.M. Satarov}

\affiliation{The Kurchatov Institute, Russian Research Centre,
123182 Moscow, Russia}

\affiliation{Institut~f\"{u}r Theoretische Physik,
J.W. Goethe Universit\"{a}t,\\
D--60054 Frankfurt am Main, Germany}

\author{W. Greiner}

\affiliation{Institut~f\"{u}r Theoretische Physik,
J.W. Goethe Universit\"{a}t,\\
D--60054 Frankfurt am Main, Germany}

\begin{abstract}
Properties of cold nuclear matter are studied within a generalized
Nambu--Jona-Lasinio model formulated on the level of constituent
nucleons.  The model parameters are chosen to reproduce simultaneously
the observed nucleon and pion masses in vacuum as well as saturation
properties of nuclear matter. The strongest constraints on these
parameters are given by the empirical values of the nucleon effective mass
and compression modulus at nuclear saturation density. A
preferable value of the cut--off momentum, determining density of active
quasinucleon states in the Dirac sea, is estimated to about 400 MeV/c.
With the most reasonable choice of model parameters 
we have found a first order phase transition of the
liquid--gas type at subsaturation densities and the gradual restoration
of chiral symmetry at about~3~times the saturation
density. Fluctuations of the scalar condensate around its mean--field
value are estimated and shown to be large in the vicinity of chiral
transition.

\end{abstract}



\maketitle

\section{Introduction}

Properties of strongly interacting matter, especially its 
possible phase transitions, are presently in the focus of
experimental and theoretical investigations.
At zero baryon density and high temperature the situation
becomes more and more clear due to continuous 
progress in lattice calculations. Latest simulations~\cite{Kar02}
including dynamical quarks with realistic masses predict 
a crossover--like deconfinement (chiral) transition 
at temperatures around 170 MeV. The situation at finite
baryon densities remains uncertain and subject to model
building. Different models predict many interesting phenomena 
in this case, ranging from restoration of chiral symmetry to color
superconductivity. 

In our recent analysis~\cite{Mis02} we tried to find 
phenomenological constraints on possible phase transitions
at high baryon densities. The main constraint comes from
the very existence of a bound state of symmetric nuclear
matter at low temperature and baryon density 
\mbox{$\rho\simeq 0.17\,{\rm fm}^{-3}$}\,. We have found
that predictions of some popular models are in direct
contradiction with this fact. 
For example, several quark models predict new phases
which have too low pressure as compared to ordinary nuclear
matter with the same baryon chemical potential. Therefore,
such phases are unstable with respect to hadronization.
Our aim in this paper is to construct a QCD motivated
effective model which is able to reproduce correctly the
nuclear saturation point and then examine its
predictions at higher densities.  

Many phenomenological models have been suggested 
to describe properties of cold nuclear matter
in terms of nucleonic degrees of freedom.
The relativistic mean--field models of the Walecka 
type~\cite{Ser85,Ser97} turned out to be rather successful
in describing properties of medium and heavy nuclei. 
One should mention also microscopic
models, dealing with nucleon--nucleon forces. Parameters of
these forces are usually extracted from scattering data and 
the observed binding energies of light nuclei 
(see e.g. Ref.~\cite{Akm98} and references therein). 
In addition to their numerical complexity, such
approaches are essentially nonrelativistic and therefore,
their accuracy diminishes at high densities.
All these models have a serious drawback, namely, they 
do not respect the chiral symmetry of strong interactions. 

There exist several chiral models which could potentially
be used for describing nuclear matter. Most popular are  
the Nambu--Jona-Lasinio (NJL) model~\cite{Nam61}
and the linear sigma model~\cite{Gel60}.  
They are able to explain spontaneous breaking 
of chiral symmetry in vacuum and its restoration at 
high energy densities. But the simplest versions of these
models fail to reproduce nuclear saturation properties.
In particular, the linear sigma model predicts only
abnormal state of nuclear matter~\cite{Lee74} where the
chiral symmetry is restored and nucleons have vanishing 
effective mass. Several more complicated models of this kind have
been suggested in Refs.~\cite{Bog83,Mis93,Pap98,Pap99}.
Although they are able to reproduce the nuclear ground
state, some new problems appear within these models. In
particular, some of them do not predict restoration of 
chiral symmetry at high baryon densities.    

There were also attempts to use the NJL model for describing 
cold nuclear matter~\mbox{\cite{Koc87,Bub96,Mis97}}\,. 
It was argued in Refs.~\cite{Koc87,Bub96} that bound nucleonic matter
with spontaneously broken chiral symmetry is not possible within 
the standard NJL. This conclusion was based on implicit assumption 
that the maximum (cut--off) momentum for constituent nucleons 
$\Lambda$ is the same  as normally used in quark models 
($\Lambda\sim 0.6$\,GeV). The authors of 
Ref.~\cite{Koc87} suggested to include 
additional (scalar--vector) interaction terms to reproduce
the observed saturation properties of nuclear matter. On the other
hand, it was shown in Ref.~\cite{Mis97} that by assuming sufficiently
low value of cutoff--momentum ($\Lambda\simeq 0.3$\,GeV) 
it is possible to achieve a bound
state at normal density even in the standard NJL model. However,
in this case the nucleon effective mass at $\rho=\rho_0$ 
is predicted by a factor of two smaller as compared 
to its empirical value.

Below we reconsider the possibility of using the NJL model
for describing cold nuclear matter. We use the generalized
version of this model including additionally the
scalar--vector interaction. We also take into account
explicit symmetry breaking effects by introducing 
a bare nucleon mass. Indeed, with this
modifications a good agreement with observed 
saturation properties of nuclear matter is achieved. 

The paper is organized as follows. In Sect.~II we formulate
the model and explain the procedure of choosing parameters.
Comparison of our results with observed characteristics 
of nuclear matter is given in Sect.~III. 
In Sect.~IV we analyze characteristics of
the chiral symmetry restoration and demonstrate the importance of 
chiral fluctuations. In Sect.~V we summarize and discuss the results. 

\section{Formulation of the model}

In this paper we return to the original formulation of 
the NJL model~\cite{Nam61} in terms of nucleonic degrees of freedom.
Namely, we consider a system of nucleons,
interacting via point--like scalar (pseudoscalar) and vector
(axial vector) 4--fermion terms. 
The corresponding Lagrangian for a spinor
nucleon field $\psi$ is written as ($\hbar=c=1$)
\begin{widetext}
\bel{lagr}
{\cal L}_0=\psib\,(i\,\dd-m_0)\,\psi+\frac{\ds G_S}{\ds 2}
\left[(\psib\psi)^2-(\mbox{$\psib\gamma_5\bm{\tau}\psi$})^2\right]
-\frac{\ds G_V}{\ds 2}\left[(\mbox{$\psib\gamma_\mu\psi$})^2+
(\mbox{$\psib\gamma_\mu\gamma_5\psi$})^2\right]\,,
\ee
\end{widetext}
where $m_0$ is the ''bare'' nucleon mass, $\bm{\tau}$ are isospin Pauli
matrices, $\gamma^\mu$ are standard Dirac matrices and
$\gamma_5=i\gamma^0\gamma^1\gamma^2\gamma^3$\,. 
$G_S$ and $G_V$ are the coupling constants for the scalar 
(pseudoscalar) and vector (axial vector) NN interactions, 
respectively. 

As will be clear later, this Lagrangian is too restrictive to reproduce
saturation properties of nucleonic matter at zero temperature.
In particular, the model should explain the phenomenological fact that
binding energy per nucleon of homogeneous isospin--symmetric matter 
is bound by about 16 MeV at the equilibrium density
$\rho=\rho_0\simeq 0.17\,{\rm fm}^{-3}$\,. 
In other words, this means that at this density the energy per baryon 
reaches its minimum value
\bel{benm}
\frac{E}{B}\raisebox{-1ex}{\LARGE
$|$}_{\rho=\rho_0}\hspace*{-1.5ex}
\simeq m_N-16\hspace*{1ex}{\rm MeV}\,,
\ee
where
$m_N=938\,{\rm MeV}$ is the nucleon mass in vacuum.

Following Refs.~\cite{Bog83,Koc87}, below we consider a generalized
version of the NJL model including an additional
8--fermion scalar--vector (SV) interaction term. The full
model Lagrangian is written as
\begin{widetext}
\bel{lagr1}
{\cal L}={\cal L}_0+\frac{\ds G_{SV}}{\ds 2}
\left[(\psib\psi)^2-(\mbox{$\psib\gamma_5\bm{\tau}\psi$})^2\right]
\left[(\mbox{$\psib\gamma_\mu\psi$})^2+
(\mbox{$\psib\gamma_\mu\gamma_5\psi$})^2\right]\,.
\ee
\end{widetext}
It is easy to see that at \mbox{$m_0= 0$} this Lagrangian
is chirally symmetric, i.e. it is invariant with respect to
$SU(2)_L\otimes SU(2)_R$ transformations. 
Below we investigate sensitivity of nuclear matter equation of state to the
coupling constants $G_S, G_V$ and $G_{SV}$\,. In fact, they 
will be chosen to fulfil the condition (\ref{benm}).

Within the mean--field (Hartree) approximation 
we replace $(\psib\Gamma\psi)^2$ operators by
\mbox{$2\hsp\psib\Gamma\psi <\psib\Gamma\psi> - <\psib\Gamma\psi>^2$}\,,
where $\Gamma$ is any combination of matrices appearing in
the interaction terms of the Lagrangian (\ref{lagr1}). The angular
brackets denote quantum--statistical averaging 
at fixed temperature $T$ and baryon density $\rho$\,. In
the considered case of isospin--symmetric matter at $T=0$ 
all terms containing $\gamma_5$ vanish and the Lagrangian takes
the form
\begin{widetext}
\bel{lagmf} {\cal L}_{\rm MFA}=
\psib\,(i\,\dd-m-\gamma^0\Sigma_V)\,\psi-U(\rho,\rho_S)\,,
\ee
\end{widetext}
where $m,\,\Sigma_V$ and $U$ are functions of scalar
($\rho_S=<\psib\psi>$) and vector ($\rho=<\psib\gamma^0\psi>$)
nucleon densities. These quantities are defined as follows
\begin{eqnarray}
&&m=m_0-\widetilde{G}_S\rho_S\,,\label{gape}\\
&&\Sigma_V=(G_V-G_{SV}\rho_S^2)\hsp\rho\,,\label{sigv}\\
&&U(\rho,\rho_S)=\frac{1}{2}\hsp (G_S\rho_S^2-G_V\rho^2+
3\hsp G_{SV}\rho^2\rho_S^2)\,.\label{upot}
\end{eqnarray}
In~\re{gape} we have introduced a renormalized scalar
coupling
\bel{rens}
\widetilde{G}_S=G_S+G_{SV}\rho^2\equiv G_S\left[1+
\alpha\left(\rho/\rho_0\right)^2\right]\,,
\ee
where $\alpha$ is a dimensionless constant, 
$\alpha=G_{SV}\rho_0^2/G_S$\,. Solutions of the gap
equation~(\ref{gape}), denoted below as $m^*_N$, have
the meaning of the nucleon effective mass.

The Euler--Lagrange equation for the fermion field $\psi$
has the form of Dirac equation with constant effective mass $m$ and
vector potential $\Sigma_V$. By using the plane wave
decomposition of $\psi$ one gets the explicit relations
\begin{eqnarray}
&&\rho_S=\nu_N\intp \frac{\ds m}{\ds\ENp}
\left[n_N(p)+n_{\ov{N}}(p)-1\right]\,,\label{sden}\\
&&\rho=\nu_N\intp\left[n_N(p)-n_{\ov{N}}(p)\right]\,,\label{vden}
\end{eqnarray}
where $\nu_N=4$ is the spin--isospin degeneracy factor for nucleons,
$n_N(p)$ and $n_{\ov{N}}(p)$ are the Fermi--Dirac occupation
numbers for nucleons and antinucleons. The last term $(-1)$ in square 
brackets of~\re{sden} originates from the negative energy levels of
the Dirac sea. It gives divergent contribution 
which needs to be regularized. Following common practice
we do this by introducing 3--dimensional
cut--off momentum $\Lambda$\,, i.e.  replacing $-1$ by
$-\Theta (\Lambda-p)$ where $\Theta(x)\equiv (1+\textrm{sgn}\hsp x)/2$\,.
One can interpret $\Lambda$ as maximum momentum determining
the number of active quasinucleon levels in the 
Dirac~sea~\footnote{
Apparently, a more realistic treatment can be achieved by introducing  
a smooth form factor in the momentum space instead of a sharp cut--off at
$p=\Lambda$~\cite{Sch94,Rip97}.}.

From Lagrangian (\ref{lagmf}) one can easily derive
the energy--momentum tensor. The energy density of homogeneous 
nuclear matter is written as
\begin{widetext}
\bel{ened}
\epsilon=\nu_N\intp\ENp\,\left[
n_N(p)+n_{\ov{N}}(p)-\Theta(\Lambda-p)\right]
+\Sigma_V\rho+U(\rho,\rho_S)+\epsilon_0\,,
\ee
\end{widetext}
where again the three--dimensional regularization
is applied to the divergent momentum integral.
The constant $\epsilon_0$ is introduced to set the energy density of the
vacuum state (\mbox{$\rho=0$}, \mbox{$m=m_N$}) equal to zero.
Obviously, energy per baryon is \mbox{$E/B\equiv\epsilon/\rho$}\,.

Below we consider isospin--symmetric matter at zero
temperature. In this case
\mbox{$n_{\ov{N}}(p)=0$} and \mbox{$n_N(p)=\Theta (p_F-p)$} where
$p_F$ is the Fermi momentum related to the baryon density,
\bel{ferm}
\rho=\frac{\ds\nu_N p_F^3}{\ds 6\pi^2}\,.
\ee
The scalar density reads
\begin{widetext}
\bel{sden1}
\rho_S=-\nu_N\int\limits^\Lambda_{p_F}\frac{d^3 p}{(2\pi)^3}
\frac{\ds m}{\ds\ENp}=\frac{\ds\nu_N m}{4\pi^2}
\left[p_F^2\Phi\left(\frac{\ds m}{\ds p_F}\right)
-\Lambda^2\Phi\left(\frac{\ds m}{\ds\Lambda}\right)\right]\,,
\ee
\end{widetext}
where
\bel{phi}
\Phi(x)=\sqrt{1+x^2}-\frac{\ds x^2}{\ds 2}
\ln{\frac{\ds\sqrt{1+x^2}+1}{\ds\sqrt{1+x^2}-1}}\,.
\ee
At sufficiently low baryon densities corresponding to $p_F<\Lambda$
the integral in the r.h.s of~\re{sden1} is positive and
$\rho_S<0$\,. However, at high densities, 
when $p_F>\Lambda$, the integral becomes negative and,
therefore, $\rho_S>0$\,. Then, according to~\re{gape}
the nucleon effective mass becomes smaller than $m_0$\,. 
Of course, at large enough momenta the approximation 
of a point--like NN interaction will fail due to the
nontrivial internal structure of nucleons. In this
situation one should explicitly consider quark degrees of
freedom. 

Expressing $\rho_S$ in terms of $m$ from \re{gape},
we rewrite~\re{ened} as
\begin{widetext}
\bel{ened1}
\epsilon=-\nu_N\int\limits^\Lambda_{p_F}\frac{d^3 p}{(2\pi)^3}\ENp
+\frac{\ds (m-m_0)^2}{\ds 2\hsp\widetilde{G}_S}+\frac{\ds
G_V\rho^2}{2} +\epsilon_0\,.
\ee
\end{widetext}
One can consider this expression as a functional of two
independent variables, $m$ and $\rho$\,. Then 
the gap equation follows from the minimization
of $\epsilon\hsp(m,\rho)$ with respect to $m$\,. 
As has been already noticed in Ref.~\cite{Koc87}, 
the inclusion of SV interaction is equivalent to
renormalizing the scalar coupling,~$G_S\to\widetilde{G}_S$\,. 

From thermodynamic identities at $T=0$, 
\bel{therm}
\mu=\frac{\ds d\epsilon}{\ds d\rho}\,,\hspace*{5mm}P=\mu\rho-\epsilon=
\rho^2\frac{\ds d}{\ds d\rho}\left(\frac{\ds\epsilon}{\ds\rho}\right)\,,
\ee
one can calculate the baryon chemical potential $\mu$ and 
pressure $P$ as functions of baryon density $\rho$\,. 
It is obvious that pressure should vanish at
the saturation point, i.e. $P(\rho_0)=0$\,.
This gives a nontrivial constraint on the model
parameters. 

Our model contains five adjustable parameters,
$\Lambda,\,m_0,\,G_S,\,G_V,\,\alpha$, which we fix by fitting the
vacuum masses of nucleons and pions, as well as 
the saturation properties of nuclear matter.
First constraint follows from the gap equation in the
vacuum. Taking \mbox{$\rho=0$},\,\,\mbox{$p_F=0,\,m=m_N$} in
Eqs.~(\ref{gape}), (\ref{rens}), (\ref{sden1}) one has
\begin{widetext}
\bel{vacm}
m_N=m_0-G_S\rho_S^{\rm vac}=m_0+\nu_NG_S\frac{\ds m_N\Lambda^2}
{4\pi^2}\Phi\left(\frac{\ds m_N}{\ds\Lambda}\right)\,.
\ee
\end{widetext}

Another constraint~\cite{Mis97} comes from the soft pion phenomenology. 
By using the Lagrangian (\ref{lagr}) in the random phase
approximation one can calculate the polarization
operator for pionic excitations (for details see Ref.~\cite{Vog91}).
By applying further the PCAC equation for the axial--vector current 
of nucleons in vacuum one gets
\bel{gorr}
m_{\pi}^2 f_{\pi}^2=m_0\hsp |\rho_S^{\rm vac}|\,,
\ee
where $m_{\pi}\simeq 140$\, MeV and $f_{\pi}\simeq 93$\, MeV are
the vacuum values of pion mass and pion decay constant, respectively.
This equation is valid in the lowest order approximation
in $m_\pi/\Lambda$ and can be considered as analogue of the 
Gell-Mann--Oakes--Renner relation~\cite{Gor68} following
from the quark structure of mesons.
According to~\re{gorr} the pure chiral limit $m_0=0$\,,
used e.g. in Ref.~\cite{Koc87}, formally corresponds to
vanishing pion mass.

Choosing different cut--off momenta $\Lambda$ as input
we use Eqs.~(\ref{vacm})--(\ref{gorr}) to find $m_0$ and~$G_S$. 
The parameters $G_V$ and~$\alpha$ can be fixed now  
by the requirement that $\epsilon/\rho$ attains the minimum value 
(\ref{benm}) at $\rho=\rho_0$\,. At fixed $\Lambda$ we first find 
$m=m^*_N (\rho)$ from the gap equation (see Eqs.~(\ref{gape}), (\ref{sden1}))
and then calculate the energy per baryon by using~\re{ened1}. The
calculation shows that at $\Lambda\lesssim 0.2$\,GeV it is not possible
to reproduce the saturation point with observable binding energy and
density at any $\alpha$\,. This becomes possible only at 
\mbox{$\Lambda\gtrsim 0.3$\,GeV}~\footnote{
Strictly speaking at \mbox{$\Lambda=0.3$\,GeV} the minimum in binding energy
appears only at \mbox{$\rho\leq 0.16\,\,{\rm fm}^{-3}$} which is
less than the $\rho_0$ value used in this work. 
However, having in mind that accuracy in $\rho_0$ is currently not
better than 10\%, we still consider such a low saturation
densities as acceptable.}.

\begin{figure}[ht]
\includegraphics[width=8.6cm]{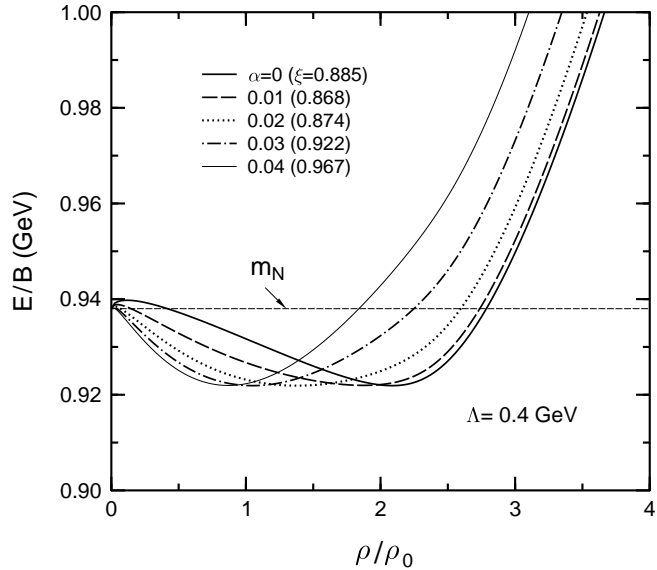}
\caption{Energy per baryon in cold nuclear matter as a function
of baryon density $\rho$\,, calculated with the cut-off
momentum $\Lambda=0.4\,\textrm{GeV}$. Different
curves corresponds to different values of scalar--vector coupling
$\alpha$\,. The parameter $\xi$ denotes the ratio of vector and scalar
coupling constants.}
\label{fig1}
\end{figure}

An illustration of the procedure used to determine parameters
$G_V$ and $\alpha$ is
given in Fig.~\ref{fig1} for the case $\Lambda=0.4$\,GeV.
At fixed $\alpha$ we choose the ratio $\xi\equiv G_V/G_S$ to 
obtain minimal energy per baryon given in~\re{benm}.
As one can see in this figure, increasing $\alpha$ shifts the
minimum in $\epsilon/\rho$ to smaller baryon densities.  In the
considered case the choice $\alpha=0$ leads to abnormal 
($m^*_N\simeq 0$) bound state at
density $\rho\simeq 2.2 \rho_0$\,.
The correct position of the saturation point is obtained
for $\alpha=0.032$\,. We repeat this analysis for other values
of~$\Lambda$ from the interval 
$0.3\,{\rm GeV}<\Lambda<0.7\,{\rm GeV}$\,. 

The corresponding parameters are listed in Table~\ref{tab1}.
The second and third columns give~$G_S$ and $m_0$ values.
determined, respectively, from Eqs.~(\ref{vacm}) and~(\ref{gorr}).
\begin{widetext}
\begin{center}
\begin{table}[ht]
\caption{Parameters of the NJL model}
\vspace*{1mm}
\label{tab1}
\begin{ruledtabular}
\begin{tabular}{ccccc}
$\Lambda$\hsp (GeV)&$G_S\hsp\textrm{(GeV\hsp fm}^3)$&
\mbox{$m_0$\hsp(MeV)}&$\alpha$&$\xi$\\
\colrule
0.3& 3.653 & 95.7 & 0 & 0.885\\
0.4& 1.677 & 41.3 & 0.032 & 0.931\\
0.5& 0.900 & 21.7 & 0.026 & 1.226\\
0.6& 0.542 & 12.9 & 0.024 & 2.224\\
0.7& 0.354 & 8.41 & 0.0225 & 4.506\\
\end{tabular}
\end{ruledtabular}
\end{table}
\end{center}
\end{widetext}
The last two columns show the
parameters $\alpha$ and $\xi$ which give the best fit of
the saturation point at fixed
cut--off momentum $\Lambda$\,. One can see from the table 
that the parameter $\alpha$
slowly decreases with~$\Lambda$ at $\Lambda\gtrsim 0.4$\,GeV. On the
other hand, the relative strength of vector interaction $\xi$
grows noticeably with increasing $\Lambda$\,. It is natural
to assume that the bare nucleon mass should not be smaller than
$3\hsp m_{0q}$\,, where $m_{0q}= (5\pm 1)$\,MeV~\cite{Pdg02}
is the isospin--averaged current mass of light quarks. From this point of view
the parameter sets with $\Lambda> 0.6$\, GeV 
should be excluded. Due to the same reason the parameter set with $m_0=0$ and 
\begin{widetext}
\bel{Koc87}
\Lambda=0.641~{\rm GeV}\,,~~G_S=0.459~{\rm GeV\hsp fm}^{-3}\,,
~~\alpha=0.023\,,~~\xi=2.92\,,
\ee
\end{widetext}
suggested in Ref.~\cite{Koc87}, does not seem reasonable.

\section{Nuclear matter within the NJL model}

\begin{figure}
\includegraphics[width=8.6cm]{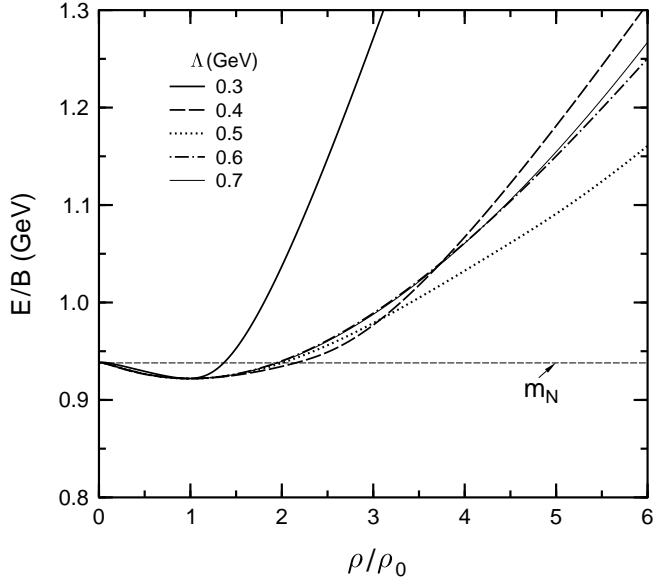}
\caption
{Energy per baryon as a function of
$\rho$ for different $\Lambda$.}
\label{fig2}
\end{figure}

Now we present the predictions of the generalized NJL model
formulated in the preceding section. Figure~\ref{fig2} shows 
density dependence of energy per baryon calculated
with parameter sets from Table~\ref{tab1}. Although the results are
similar at $\rho\lesssim\rho_0$\,, at higher densities 
one can see strong sensitivity to $\Lambda$. The curve for
$\Lambda=0.3$\,GeV shows especially large deviation from the others.
As already noted  above,
one should be very cautious by applying the NJL model at high densities,
corresponding to $p_F>\Lambda$\,. This is particularly important 
for sets with smaller $\Lambda$\,. Indeed, for $\Lambda=0.3$\,GeV
$p_F>\Lambda$  holds already at $\rho>1.4\hsp\rho_0$
\footnote{
For $\Lambda=0.4$~and $0.5$\,GeV this takes place 
at $3.3\hsp\rho_0$ and $6.5\hsp\rho_0$\,,
respectively. For $\Lambda\gtrsim 0.6$\,GeV the maximal 
densities become larger than $10\hsp \rho_0$. Nucleons will
be completely dissolved at such high densities.}.

\begin{figure}
\includegraphics[width=8.6cm]{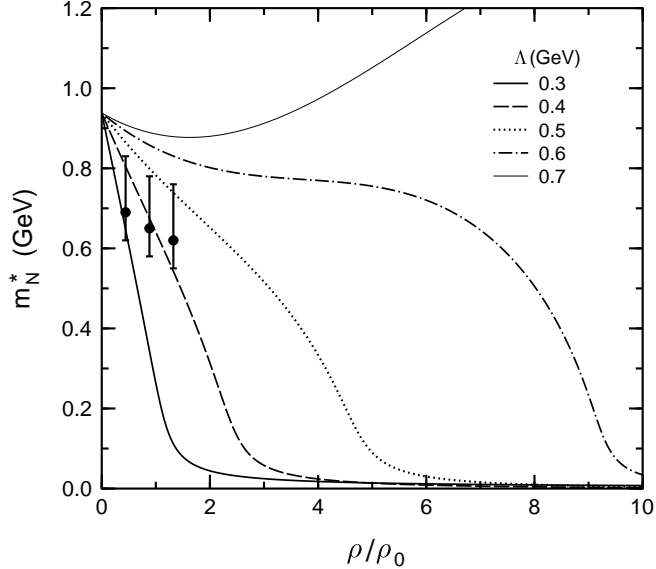}
\caption
{Nucleon effective mass as a function of
baryon density for different $\Lambda$.
The dots show estimates~\cite{Fur96}, based on the QCD sum rules.}
\label{fig3}
\end{figure}

Figure~\ref{fig3} shows density dependence of the nucleon effective mass 
$m_N^*$ calculated with the same parameter sets.
On can see that the $m_N^*$ behaviour depends strongly on $\Lambda$ 
even at low densities. For example, at $\Lambda=0.7$\,GeV
it has a very peculiar behavior. In this case $m_N^*$  
first increases, reaching maximum value
$m_N^*\simeq 1.4$\,GeV at $\rho\simeq 10\hsp\rho_0$\,, and then drops
down approaching the bare mass at $\rho\sim 17\hsp\rho_0$. 
For all $\Lambda$ the nucleon effective mass becomes small
at sufficiently high densities, but it does not vanish. This would be 
possible only in the case of exact chiral symmetry ($m_0=0$)
when $m_N^*=0$ is an exact solution of the gap equation. Therefore, 
for all parameter sets with $m_0\neq 0$ the chiral symmetry 
is only approximately restored at high densities.
By filled dots in Fig.~\ref{fig3} we show the mass values 
obtained from the QCD sum rules~\cite{Fur96}. Although the error bars are
large, these estimates clearly favour the parameter set with 
$\Lambda=0.4$\,GeV. As compared to vacuum, a 30\% reduction of 
the nucleon mass at $\rho=\rho_0$ is predicted in this case.
The results for different $\Lambda$ are presented in 
Table~\ref{tab2}.

\begin{widetext}
\begin{center}
\begin{table}[ht]
\caption{Properties of nuclear matter within the NJL model}
\vspace*{1mm}
\label{tab2}
\begin{ruledtabular}
\begin{tabular}{ccccc}
$\Lambda$\hsp (GeV)&$m_N^*(\rho_0)/m_N$&
$K(\rho_0)$\hsp (MeV)&$\Sigma_{\pi N}\hsp ({\rm MeV})$
&$\rho_c/\rho_0$\\
\colrule
0.3& 0.385 & 1350 & 101  &  1.24 \\
0.4& 0.683 & 285  & 42.1 &  2.66 \\
0.5& 0.834 & 322  & 25.1 &  5.08 \\
0.6& 0.905 & 347  & 15.9 &  9.51 \\
0.7& 0.944 & 337  & 10.9 &  16.3 \\
\end{tabular}
\end{ruledtabular}
\end{table}
\end{center}
\end{widetext}

\begin{figure}
\includegraphics[width=8.6cm]{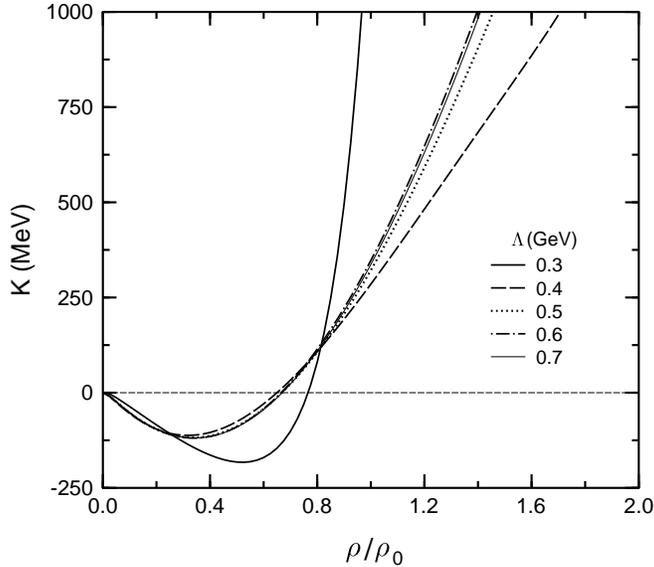}
\caption
{Compression modulus as a function of baryon density
at different $\Lambda$.}
\label{fig4}
\end{figure}

The compression modulus of nuclear matter, $K$, is 
an important characteristics of the equation of state.
It is defined as  
\begin{widetext}
\bel{comm}
\frac{\ds K}{\ds 9}=\frac{\ds dP}{\ds d\rho}=
\rho\frac{\ds d\mu}{\ds d\rho}=
\rho\frac{\ds d^2\epsilon}{\ds d\hsp\rho^2}\,,
\ee
\end{widetext}
where the second and third equalities follow 
from~\re{therm}. The density dependence of $K$ is
shown in Fig.~\ref{fig4} for different values of
$\Lambda$\,. Negative $K$ values correspond to
mechanical instability of matter with respect to density
fluctuations. The density interval with $K<0$
is known as the spinodal region of a first order (liquid--gas) 
phase transition. As one can see from Fig.~\ref{fig4}, this
phase transition is undoubtedly predicted in all cases
considered. This is not surprising because such a phase
transition is a consequence of the saturating
property of the equation of state. Indeed, since pressure
is zero at $\rho=0$ and $\rho=\rho_0$ is must be
a nonmonotonous function of $\rho$\,.    
Fig.~\ref{fig4} also shows that the compression modulus  
changes very rapidly in the vicinity of the saturation
point, especially in the case of $\Lambda=0.3$\,GeV.
The values of $K$ at $\rho=\rho_0$  
are given in Table~\ref{tab2}. One can see that
the parameter set with  $\Lambda=0.4$\,GeV gives the 
$K$ value which is close to the phenomenological estimate,
$K\simeq 250-270$\,MeV~\cite{Ma01}.   

\begin{figure}
\includegraphics[width=8.6cm]{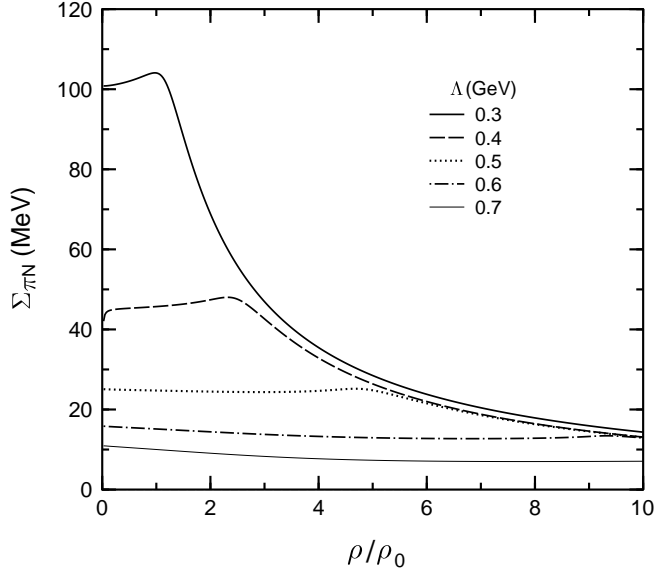}
\caption
{Pion--nucleon sigma term as a function of baryon density
at different $\Lambda$.}
\label{fig5}
\end{figure}

In studies of in--medium effects,
the so--called pion--nucleon sigma term, 
$\Sigma_{\pi N}$\,, is often used~\cite{Vog91,Coh92}.   
At small densities it can be defined as
\bel{sigt}
\Sigma_{\pi N}=\left(1-\frac{\ds <\ov{q}q>}
{\ds ~~<\ov{q}q>_{\rm vac}}\right)
\frac{\ds m_\pi^2f_\pi^2}{\rho}\,,
\ee
where $<\ov{q}q>$ is the quark condensate or, in the
context of our model the quark scalar density. 
Assuming that the scalar nucleon and quark densities are proportional to 
each other, we calculate $\Sigma_{\pi N}$ by
using~\re{sigt} with the replacement 
\mbox{$<\ov{q}q>/<\ov{q}q>_{\rm vac}\to\rho_S/\rho_S^{\rm vac}$}.
The results of this calculation are shown in Fig.~\ref{fig5}.
The $\Sigma_{\pi N}$ values at $\rho\to 0$ are also given in
Table~\ref{tab2}. At $\Lambda=0.4$\,GeV the calculated
value is close to the estimate 
$\Sigma_{\pi N}\simeq (35\pm 5)$\,MeV, obtained
from the chiral perturbation theory~\cite{Gas91}.   

\begin{figure}
\includegraphics[width=8.6cm]{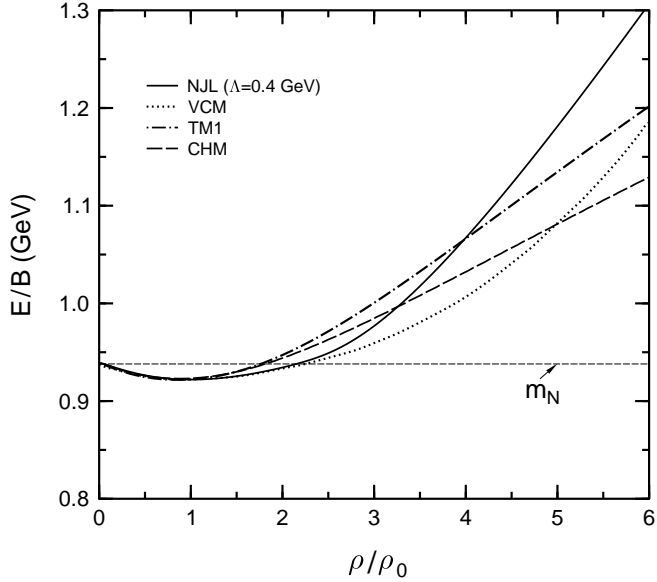}
\caption
{Energy per baryon calculated within different
nuclear models (see text).}
\label{fig6}
\end{figure}
 

Based on the results presented above we come to the
conclusion that the generalized NJL model with $\Lambda\simeq 0.4$\,GeV
gives the best description of cold nuclear matter in
comparison with other choices of $\Lambda$\,. In Fig.~\ref{fig6} we compare energies
per baryon predicted by our model with the results of other
nuclear models namely, the Walecka--type TM1 model~\cite{Sug94}, the 
Chiral Hadron Model (CHM)~\cite{Pap99} and the 
Variational Chain Model (VCM)~\cite{Akm98}. Note that the 
VCM includes two-- and three--body NN forces and takes into
account the correlation effects. One can see that at 
densities $\rho\lesssim 2\,\rho_0$ our model 
and other models give very similar results. Thus, with
$\Lambda= 0.4$\,GeV our fit of nuclear saturation
properties is at least of the same quality as in other
models listed above.             
     
\section{Identification of the chiral transition}

\begin{figure}
\includegraphics[width=8.6cm]{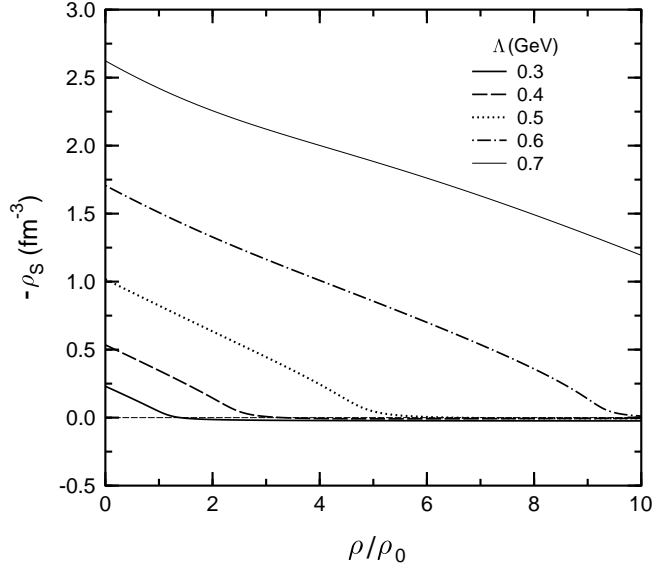}
\caption
{Scalar condensate density as a function of
baryon density for different $\Lambda$.}
\label{fig7}
\end{figure}

In this section we examine properties and possible
signatures of the chiral transition in cold nuclear
matter. As noted in Ref.~\cite{Wel82} the direct signature of
the chiral symmetry restoration is not the effective mass (the
latter may even increase with energy density), but the
vanishing chiral condensate. In our approach the chiral
condensate coincides with the scalar nucleon density 
$\rho_S=<\psib\psi>$\,. Figure~\ref{fig7} displays
this quantity calculated for the same parameter sets
as before. Unlike the effective mass, 
$|\rho_S|$ decreases practically linearly with~$\rho$ 
at not too high densities. We define the critical 
baryon density of the chiral transition,~$\rho_c$, by the linear 
extrapolation of $\rho_S(\rho)$ to zero. Numerical values of these 
densities (in units of~$\rho_0$) are given in the last column 
of Table~\ref{tab2}.

\begin{figure}
\includegraphics[width=8.6cm]{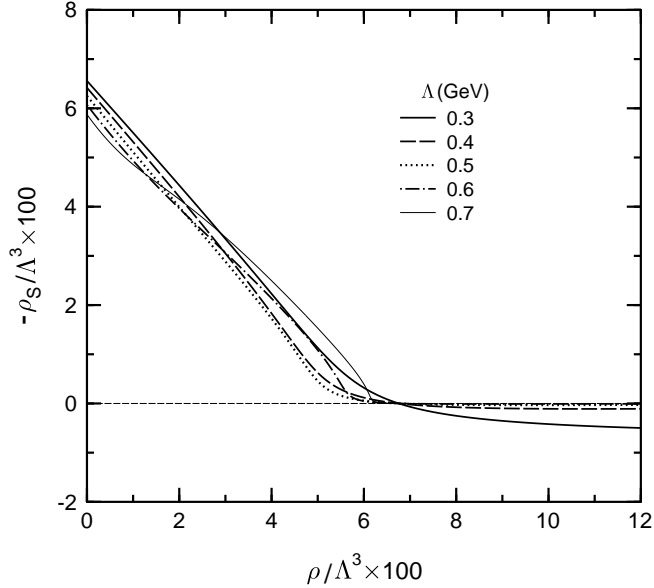}
\caption
{Same as Fig.~\ref{fig7}, but with scaled densities.}
\label{fig8}
\end{figure}   

The results of Fig.~\ref{fig7} suggest that scalar
density curves for different $\Lambda$ have similar behavior 
and roughly can be obtained
from one another by rescaling the axes.
In Fig.~\ref{fig8} we show the same results, but 
scalar and vector nucleon densities are now scaled by 
the factor~$\Lambda^{-3}$\,. 
One can see that in these variables the results for different 
$\Lambda$ are indeed very close to each other. It is evident now 
that the critical density for the chiral transition 
may be roughly estimated~as
\bel{crdp}
\rho_c\simeq 0.06\hsp\Lambda^3\,.
\ee
The corresponding Fermi momentum $p_F$ at
$\rho\simeq\rho_c$ equals approximately $0.96\hsp\Lambda$\,. Therefore, 
the region of the chiral transition is close to the boundary of the model
applicability (see discussion above). 

It is instructive to compare the vacuum value of nucleon scalar
density with the quark condensate $<\ov{q}q>$\,. 
The phenomenological value of the quark condensate in vacuum is~\cite{Coh92}
\bel{qcde}
|<\ov{q}q>|_{\rm vac}\simeq 2\,|<\ov{u}u>|_{\rm vac}
\simeq 2\,(225\pm 25\,{\rm MeV})^3\simeq (3\pm 1)\,{\rm fm}^{-3}\,.
\ee
According to Fig.~\ref{fig7}, for $\Lambda=0.4$\,GeV 
one gets the value \mbox{$|\rho_S^{\rm vac}|
\simeq 0.54\,{\rm fm}^{-3}$}\, which is by about factor 6
smaller than \mbox{$|<\ov{q}q>|_{\rm vac}$}. For $\Lambda=0.5$\,GeV 
the ratio is close to 1/3 which might be expected from
the naive constituent quark model. For higher $\Lambda$
the predicted nucleon condensate is comparable with
\mbox{$<\ov{q}q>_{\rm vac}$}\,.

\begin{figure}
\vspace*{-4cm}
\includegraphics[width=10cm]{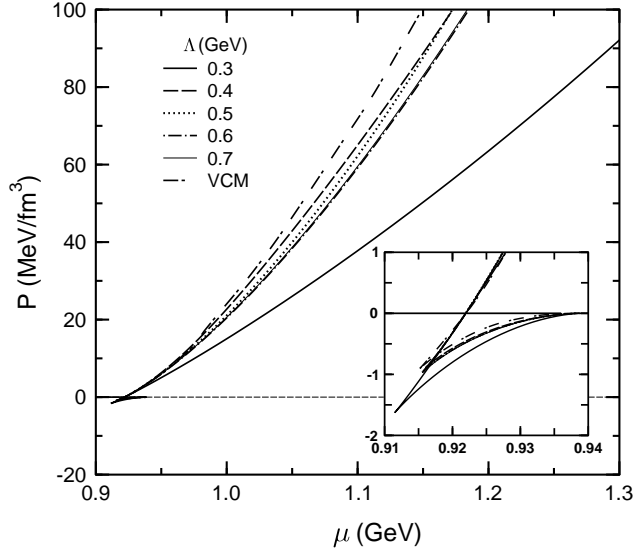}
\vspace*{-4cm}     
\caption
{Pressure as a function of chemical potential for different
$\Lambda$\,. The curve labelled by VCM is calculated within
the model of Ref.~\cite{Akm98}. The insert shows $P(\mu)$ 
in the region of liquid--gas phase transition.} 
\label{fig9}
\end{figure}

When looking for possible phase transitions 
it is most useful to calculate 
pressure $P$ as a function of chemical potential $\mu$\,.
A first order phase transition is signalled by 
appearance of several branches of $P(\mu)$\,.
At fixed $\mu$ only the branch with highest
pressure corresponds to a stable phase. The point of phase
transition, $\mu=\mu_c$\,, is given by intersection
of a stable and a metastable branches. At this point 
$dP/d\mu$ exhibits a jump which corresponds to
difference of densities between coexisting phases. 
Figure~\ref{fig9} shows the
pressure curves predicted by our model for different
$\Lambda$\,. At small $\mu$ one can see the presence of a 
first--order phase transition which takes place 
at $\mu\simeq m_N-16\,{\rm MeV}\simeq 922$\,MeV for all considered 
sets of model parameters. The region of mixed phase
covers subnuclear densities, $\rho<\rho_0$\,. This
is a liquid--gas phase transition which is well--known  
in nuclear physics. It is responsible for the
multifragmentation phenomenon in hot nuclear systems. 
There is no clear evidence of other 
non--trivial behaviors of $P(\mu)$\,. Calculating
$P$ as a function of $\rho$ shows
that pressure varies rather smoothly at 
$\rho\simeq\rho_c$\,\,i.e. in the vicinity of the 
chiral transition. 

\begin{figure}[ht]
\vspace*{-7cm}\includegraphics[width=10cm]{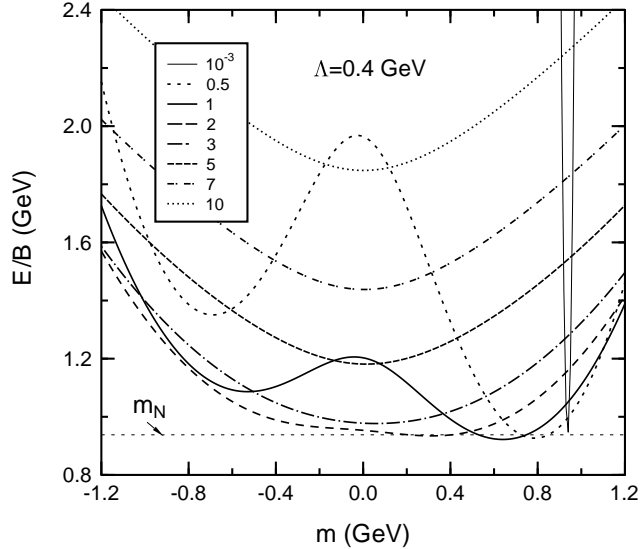}
\caption
{Energy per baryon as a function of
nucleon mass at fixed baryon densities (shown in units
of $\rho_0$ in the key box) within the NJL model with $\Lambda=0.4$\,GeV.}
\label{fig10}
\end{figure}
The results presented above were obtained assuming validity
of the men--field approximation. In particular, 
fluctuations of the scalar condensate $\rho_S$ or,
equivalently, the nucleon effective mass $m$ 
around their values determined from the gap
equation were completely disregarded. As we shall see below, this
approximation becomes rather poor at $\rho\sim\rho_c$\,.
To estimate possible magnitude of mass fluctuations,    
it is instructive to study the energy per baryon, $\epsilon/\rho$, 
as a functional of $\rho$ and $m$, i.e. without using the gap equation. 
By fixing baryon density, one obtains 
the profiles of $\epsilon/\rho$ as functions of $m$\,. 
Their minima correspond to $m=m^*_N(\rho)$\,.
The results of calculations are shown in Fig.~\ref{fig10} 
for the parameter set with \mbox{$\Lambda=0.4$\,GeV}. 

As seen in Fig.~\ref{fig10}, at $\rho\to 0$ energy per
baryon is a sharp function of mass near its minimum at 
$m\simeq m_N$\,. As $\rho$ grows, the minimum of 
$\epsilon/\rho$ shifts to smaller masses
and its width significantly increases.
At $\rho=\rho_0$ the minimum corresponds to the saturation
point where $m=m^*_N\simeq 0.683\,m_N$. When $\rho$
approaches the characteristic density of chiral transition
$\rho\simeq\rho_c$ the minimum shifts to $m\simeq 0$ and 
$\epsilon/\rho$ curves become rather flat. For example, at
$\rho=2\hsp\rho_0$ the width of mass distribution becomes  
comparable with the value $m_N^*\simeq 0.3$\,GeV
obtained from the gap equation. At higher densities 
widths of $\epsilon/\rho$ remain approximately the same. 

\begin{figure}
\vspace*{-4cm}
\includegraphics[width=10cm]{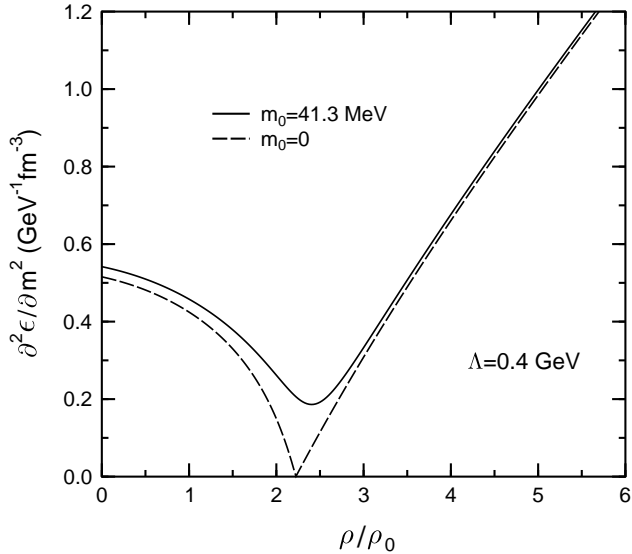}
\vspace*{-4cm}
\caption
{Second derivative of energy density with respect to
nucleon effective mass as a function of baryon density.
Solid line represents the results of calculation
with the parameters from second line of Table~\ref{tab1}.
Dashed line corresponds to the same calculation, but
with $m_0=0$\,.}
\label{fig11}
\end{figure}

To qualitatively characterize this behavior, we
evaluate $\partial^2\epsilon/\partial m^2$, the second derivative 
of $\epsilon$ with respect to $m$ at fixed $\rho$ and 
$m=m^*_N(\rho)$. This quantity describes the stiffness
of the effective potential along the mass coordinate. 
It is analogous to the compression modulus characterizing
the stiffness with respect to the density fluctuation. 
The results are shown in Fig.~\ref{fig11} by the solid line.
For comparison, the dashed line shows the same
calculation, but with vanishing bare mass $m_0$\,.
One can clearly see that the stiffness drops  
dramatically at the point of chiral symmetry restoration, 
where $m^*_N$ and $\rho_S$ become small. 
In the case $m_0=0$ the stiffness vanishes at $\rho=\rho_c$ 
and, therefore, the fluctuations diverge. Such behavior is
expected for a second order phase transition. 
It is well known that the
mean--field approximation breaks down in the vicinity of
the critical point. However, in a realistic case 
of $m_0\neq 0$ one can speak only about a significant
enhancement of fluctuations near the chiral transition
point.

\begin{figure}
\includegraphics[width=8.6cm]{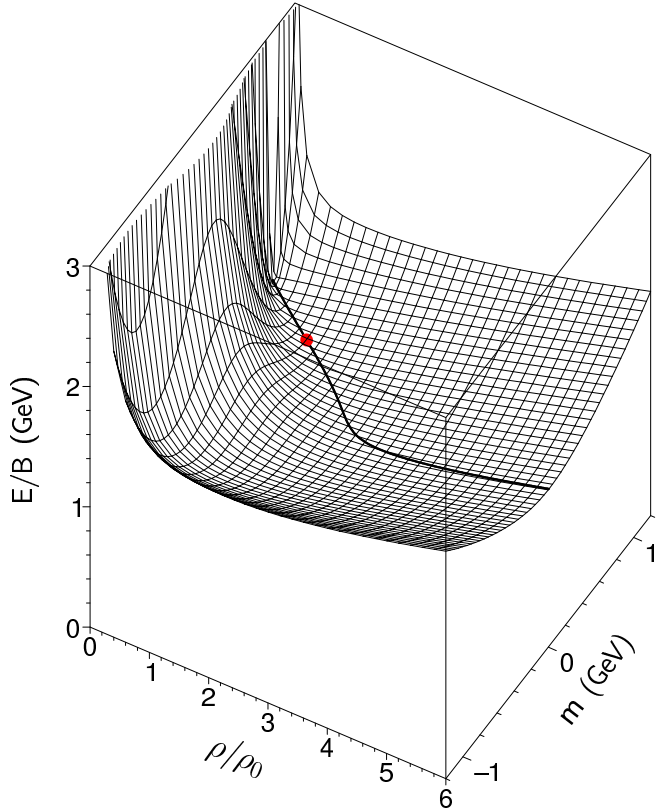}
\caption
{Energy functional in the coordinates ''effective
mass--baryon density'' calculated within the NJL model
with $\Lambda=0.4$\,GeV. Thick line shows 
local minima given by the gap equation. The dot marks the
global minimum corresponding to the nuclear saturation
point.}
\label{fig12}
\end{figure}

Finally, in Fig.~\ref{fig12} we present a three--dimensional
plot of $\epsilon/\rho$ as a function of $m$ and $\rho$\,,  
calculated for the same set with $\Lambda=0.4$\,GeV. 
Thick line connecting local minima of this surface
corresponds to solutions of the gap equation~(\ref{gape}).  
One can easily notice approximate symmetry of $\epsilon/\rho$
with respect to reflections $m\to -m$\,. At low densities
one sees two local minima~\footnote{
More careful analysis shows that minima at negative $m$
are in fact saddle points in \mbox{4-dimensional} chiral space.}
separated by a barrier (see also Fig.~\ref{fig10}).
At $\rho\simeq\rho_c$ the barrier disappears and 
the surface becomes rather flat along the mass axis. 

\begin{figure}
\includegraphics[width=8.6cm]{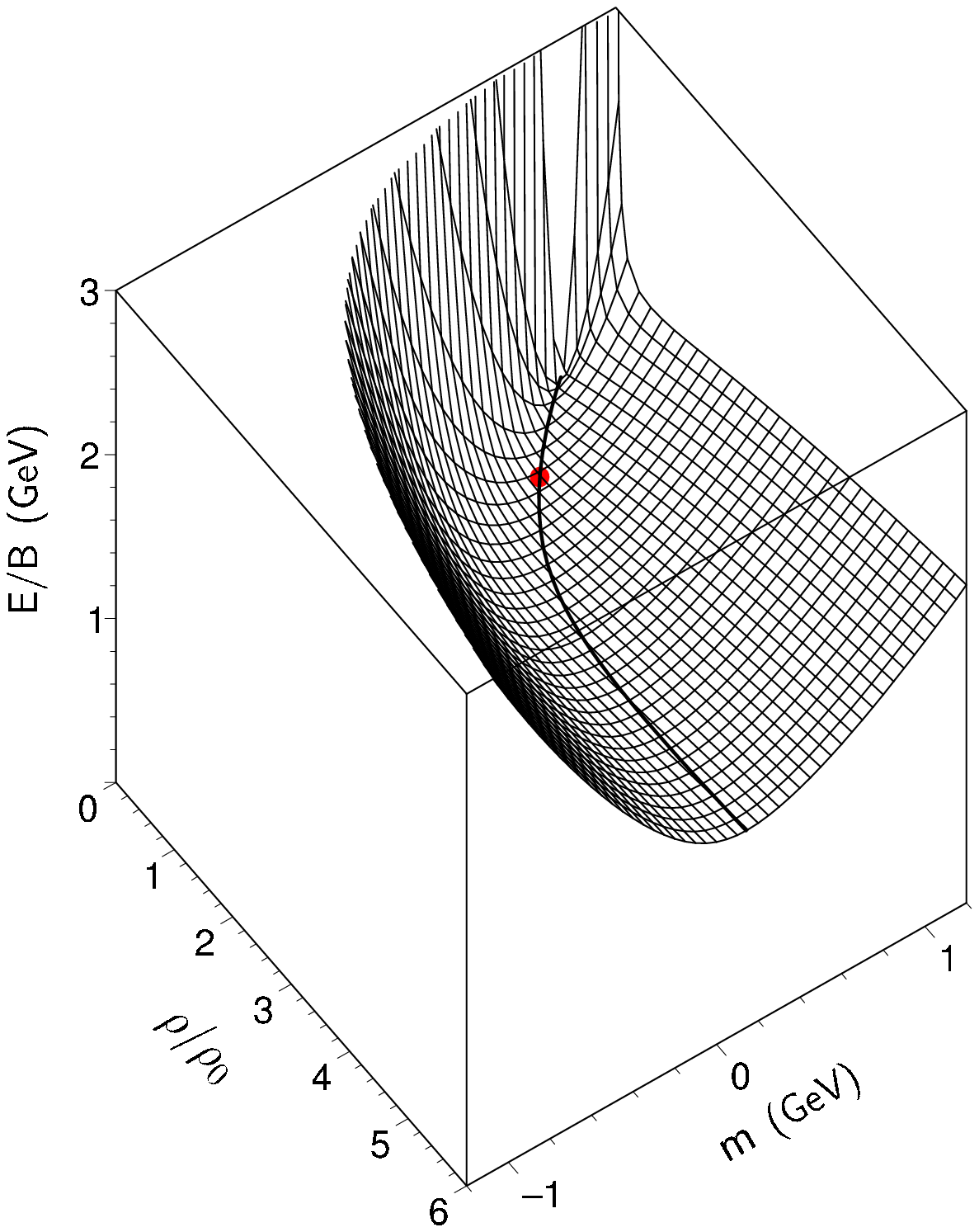}
\caption
{Same as Fig.~\ref{fig12}, but
calculated within the TM1 model. 
}
\label{fig13}
\end{figure}

It is interesting to note that in the case of exact chiral 
symmetry ($m_0=0$) the energy functional would be an even
function of the effective mass $m$\,, i.e.
$\epsilon\hsp (-m,\rho)=\epsilon\hsp (m,\rho)$\,.
In particular, at $\rho\to 0$ it has two degenerate minima
at $m=\pm\hsp m_N$\,. This symmetry is simply a particular case
of the chiral rotation corresponding to angle $\pi$\,. By using this
fact one can prove the following theorem, valid also at
$T\neq 0$\,. Namely, the thermodynamic potential of any 
chirally--symmetric theory, $\Omega =-P (m,T,\mu)$
is an even function of the fermion mass $m$ 
and has an extremum (maximum, minimum or saddle point) at
$m=0$\,. Explicit symmetry breaking terms make $\Omega$
slightly tilted towards $m>0$\,. All known chiral models 
obey this theorem. 
In contrast, traditional nuclear models of Walecka type
strongly violate it. They assign no special 
significance to the point $m=0$\,. Such models are tuned
to the saturation point of nuclear matter and can not be
extrapolated far away from this point. As an example, in
Fig.~\ref{fig13} we show the energy surface predicted by
the TM1 version of the relativistic mean--field
model~\cite{Sug94}. One can see that the model does not
reveal any symmetry with respect to $m\to -m$\,.
Moreover the minima at $m<0$ do not appear at all.       

\section{Summary and discussion}

In this paper we describe cold nuclear matter using
a generalized version of the NJL model including additional
scalar--vector interaction terms. 
We formulate the model in terms of nucleonic degrees
of freedom bearing in mind that normal nuclei are indeed
made of nucleons. We have demonstrated that 
this model is able to reproduce well 
observed saturation properties of nuclear matter such as
equilibrium density, binding energy, compression modulus
and nucleon effective mass at $\rho=\rho_0$\,.The best fit
of nuclear properties is achieved with the
cut--off momentum $\Lambda\simeq 0.4$\,GeV 
which is noticeably smaller than usually assumed for 
quark--based models. This is an indication that using nucleonic
quasiparticles is only justified at low momenta $p<\Lambda$\,.
At higher $p$ the quark degrees of freedom should be
included explicitly. 

The model predicts two interesting features. 
First, it reveals a first order phase transition
of the liquid--gas type occurring at subsaturation densities.
This phase transition automatically follows from the
existence of the nuclear bound state and, therefore, must
be present in any realistic model of nuclear matter.
Second, the model predicts an approximate restoration of 
chiral symmetry at high baryon densities, $\rho\gtrsim 3\rho_0$\,.
This feature is also quite natural for the NJL model and
is expected in any chiral model. But as follows from
our analysis, this chiral transition does not produce
any significant peculiarities in thermodynamical
quantities. Only in the case of exact chiral symmetry ($m_0=0$) 
this phenomenon would correspond
to a true second order phase transition. It is demonstrated
that the chiral transition is manifested by increased
fluctuations of the order parameter, i.e. the scalar
condensate. 

We want to stress that our model
is able to describe simultaneously the
saturation properties of nuclear matter and restoration of 
chiral symmetry at high baryon densities. The model exhibits 
only one first order phase transition of the liquid--gas
type at subsaturation densities. Restoration of chiral
symmetry develops gradually with increasing baryon density
and does not lead to any phase transition.
This confirms conclusions of Ref.~\cite{Mis02}
where we have shown that some popular predictions
of the first order chiral phase transition are incompatible 
with phenomenological constraints.

In conclusion, we have achieved a good description of cold
nuclear matter on the basis of a 
relatively simple chiral model. Within this model
normal nuclei are interpreted as droplets of baryon--rich 
matter in a phase with spontaneously broken chiral symmetry. The gradual
approach to a phase with restored chiral symmetry is
predicted at baryon densities above~$3\rho_0$\,.
This model can be easily extended to nonzero
temperatures and finite nuclei. 

Finally we want to emphasize the following.
The liquid--gas phase transition discussed above
can be observed via multiple production
of nuclear fragments in nuclear reactions at intermediate
energies. Nuclear multifragmentation is well
established phenomenon and many observations reveal trends
expected for such a phase transition. 
The situation with the chiral/deconfinement transition 
in baryon--rich matter is less certain.
Despite of many interesting suggestions a similar unique phenomenon
linked to this transition has not been established
yet. We believe that effects of the chiral symmetry restoration 
will be most clearly seen in nuclear collisions at energies
of a few 10 AGeV, when highest baryon densities are
expected. But to find them one should be prepared for a
thorough study of observables sensitive to chiral
fluctuations.          

\begin{acknowledgments}
This work has been supported by the DFG Grant 436 RUS 113/711/0-1
and the RFBR Grants 00--15--96590, 03-02-04007.
\end{acknowledgments}

\end{document}